# Metal to Mott Insulator Transition in Two-dimensional 1T-TaSe$_2$

Ning Tian[1,2,3,4,5,†], Zhe Huang[6,†], Bo Gyu Jang[7,†], Shuaifei Guo[1,2,3,4,5], Ya-Jun Yan[8], Jingjing Gao[1,2,3,4,5], Yijun Yu[1,2,3,4,5], Jinwoong Hwang[9], Meixiao Wang[10,11], Xuan Luo[12], Yu Ping Sun[12,13,14], Zhongkai Liu[10,11], Dong-Lai Feng[8], Xianhui Chen[15], Sung-Kwan Mo[9], Minjae Kim[7], Young-Woo Son[7,*], Dawei Shen[6,*], Wei Ruan[1,3,4,5,*] and Yuanbo Zhang[1,2,3,4,5,*]

[1]*State Key Laboratory of Surface Physics and Department of Physics, Fudan University, Shanghai 200438, China*
[2]*Shanghai Qi Zhi Institute, Shanghai 200232, China*
[3]*Shanghai Research Center for Quantum Sciences, Shanghai 201315, China*
[4]*Institute for Nanoelectronic Devices and Quantum Computing, Fudan University, Shanghai 200433, China*
[5]*Zhangjiang Fudan International Innovation Center, Fudan University, Shanghai 201210, China*
[6]*State Key Laboratory of Functional Materials for Informatics, Shanghai Institute of Microsystem and Information Technology (SIMIT), Chinese Academy of Sciences, Shanghai, 200050, China*
[7]*Korea Institute for Advanced study, Seoul 02455, Korea*
[8]*School of Emerging Technology and Department of Physics, University of Science and Technology of China, Hefei 230026, China*
[9]*Advanced Light Source, Lawrence Berkeley National Laboratory, Berkeley, CA 94720, USA.*
[10]*School of Physical Science and Technology, ShanghaiTech University, Shanghai 201210, China*
[11]*ShanghaiTech Laboratory for Topological Physics, Shanghai 200031, China*
[12]*Key Laboratory of Materials Physics, Institute of Solid State Physics, HFIPS, Chinese Academy of Sciences, Hefei 230031, China*
[13]*High Magnetic Field Laboratory, HFIPS, Chinese Academy of Science, Hefei 230031, China*
[14]*Collaborative Innovation Centre of Advanced Microstructures, Nanjing University, Nanjing 210093, China*
[15]*Department of Physics, University of Science and Technology of China, and Key Laboratory of Strongly Coupled Quantum Matter Physics, Chinese Academy of Science, Hefei 230026, China*

[†]These authors contributed to this work
[*]Correspondence should be addressed to Y.Z. (zhyb@fudan.edu.cn), W.R. (weiruan@fudan.edu.cn), D.S. (dwshen@mail.sim.ac.cn) and Y.-W.S. (hand@kias.re.kr)

**When electron-electron interaction dominates over other electronic energy scales, exotic, collective phenomena often emerge out of seemingly ordinary matter. The strongly correlated phenomena, such as quantum spin liquid[1,2] and unconventional superconductivity[3], represent a major research frontier and a constant source of inspiration[4,5]. Central to strongly correlated physics is the concept of Mott insulator, from which various other correlated phases derive[6,7]. The advent of two-dimensional (2D) materials brings unprecedented opportunities to the study of strongly correlated physics in the 2D limit[8]. In particular, the enhanced correlation and extreme tunability of 2D materials enables exploring strongly correlated systems across uncharted parameter space[9–14]. Here, we discover an intriguing metal to Mott insulator transition in 1T-TaSe$_2$ as the material is thinned down to atomic thicknesses. Specifically, we discover, for the first time, that the bulk metallicity of 1T-TaSe$_2$ arises from a band crossing Fermi level. Reducing the dimensionality effectively quenches the kinetic energy of the initially itinerant electrons, and drives the material into a Mott insulating state. The dimensionality-driven Metal to Mott insulator transition resolves the long-standing dichotomy between metallic bulk and insulating surface of 1T-TaSe$_2$. Our results additionally establish 1T-TaSe$_2$ as an ideal variable system for exploring various strongly correlated phenomena.**

Dimensionality plays a fundamental role in condensed matter physics. The reduction in dimensionality fundamentally alters the electronic structure of a material, often with profound consequences[15–17]. This is best exemplified by the myriad of novel phenomena discovered in 2D materials, especially when the material family expands into major new branches of condensed matter research[12,18,19]. The emerging 2D strongly correlated materials bring new opportunities—the lack of screening in the third dimension dramatically enhances the electron-electron interaction in 2D[9–11,20]. The much-enhanced interaction enriches the already vast variety of novel electronic structures in 2D materials. Specifically, the many-body interaction (characterized by the on-site Coulomb energy $U$) offers a new tuning knob to the diverse single-particle physics in 2D materials that largely originates from the inter-site hopping of electrons (characterized by the width of the resulting energy band $W$). The competition/cooperation of the two energy scales may lead to unexpected quantum phenomena.

Correlated effects exist in various layered van der Waals crystals[8,21]. When thinned down to atomic thicknesses, these 2D crystals represent the thinnest possible materials that lend themselves to external modulations such as gate doping[22–25]. Recent discovery of moiré superlattices in van der Waals heterostructures has taken such tunability to unprecedented levels. For example, it has been demonstrated that both carrier doping and bandwidth $W$ can be readily modulated by gate electric field[26,27]. Such tunability stems from the large moiré unit cell (consisting of thousands of atoms) that effectively reduces the electric field/doping required for modulating the correlated physics in the heterostructure[28]. The wide tunability comes, however, at a price—the large unit cell reduces both correlation energy scale $U$ and bandwidth $W$ to the order of ~ 10 meV (ref. [27-29]). Further exploring rich strongly correlated phenomena in 2D, as well as their potential applications, calls for fully-tunable material systems that retain a large energy scale.

Among correlated 2D materials, transition metal dichalcogenides 1T-TaX$_2$ (X=S, Se) distinguish themselves as a potentially fully-tunable material system with a high correlation energy. The ground state of the materials features a peculiar star-of-David charge density wave (CDW) superlattice[30,31]. Each star-of-David unit cell contains 13 $5d$ conduction electrons contributed by the 13 Ta atoms in the cluster [Fig. 2b]. The formation of the triangular superlattice, however, localizes all but one conduction electron on each cluster. The small star-of-David unit cell (compared to moiré unit cells in typical 2D heterostructure) dictates that the Coulomb energy of the lone electron remain high (up to ~ 500 meV; ref. [32–34]). Meanwhile, the unit cell is still large enough to ensure that the materials stay tunable—a charge doping of $7.3 \times 10^{13}$ cm$^2$ can fill/deplete the entire conduction band; this level of modulation is well within reach with ionic gating[35]. Indeed, signs of strong correlation have been observed in monolayer 1T-TaX$_2$ and their heterostructures[36–38]. There are, however, still controversies on the nature of the insulating states in 1T-TaS$_2$ (ref. [39–42]). The issue is even more severe in 1T-TaSe$_2$: bulk 1T-TaSe$_2$ is metallic, but spectroscopic measurements of the material, either in bulk or monolayer form, always produce a large gap at Fermi level[34,37,43]. These problems represent a crucial missing piece to the strong correlation puzzle in 1T-TaX$_2$. Their solution may provide important insights into 2D strongly correlated many-body physics in general.

Here, we directly address these open problems by probing the evolution of the electronic structure in 1T-TaSe$_2$ as the dimensionality of the material is continuously lowered. At the three-dimensional (3D) bulk limit, we discover a dispersive band crossing Fermi level, which has not been observed previously and thus solves the long running mystery of bulk metallicity in 1T-TaSe$_2$. Meanwhile, at the 2D limit, we find unequivocal evidence that 1T-TaSe$_2$ is, indeed, a Mott insulator. We further identify a metal to Mott insulator transition at a critical thickness of seven layers. Detailed analysis of angle-resolved photoemission spectroscopy (ARPES) and scanning tunneling spectroscopy (STS) data corroborated by first principles calculations reveals that the transition is driven by dimensionality crossover: lowering the dimensionality effectively reduces the width $W$ of the dispersive band, and induces the transition when Coulomb energy $U$ dominates over $W$. The dimensionality crossover is also at work on the surface of a bulk metal. The mechanism makes 1T-TaSe$_2$ surface layers a Mott insulator, and resolves the dichotomy between metallic bulk and insulating surface of 1T-TaSe$_2$. These results establish 1T-TaSe$_2$ as a fully tunable, correlated 2D material with a high correlation energy.

We start with electronic transport characterization of 1T-TaSe$_2$ as the sample thickness is varied from bulk down to monolayer. Thin flakes of 1T-TaSe$_2$ are mechanically exfoliated on the Si substrates covered with 285 nm SiO$_2$. The optical image of a representative few-layer 1T-TaSe$_2$ flake is shown in Fig. 1a. We identify the number of layers from the optical contrast (Fig. 1a, c) in combination with atomic force microscopy (AFM) measurement (Fig. 1b, c). Metal contacts are then defined on the flakes—by direct deposition of Cr and Au, typically 3 nm and 60 nm, respectively, through stencil masks—for subsequent transport measurements. Figure 1d displays the sheet resistance, $R_\square$, measured as a function of temperature, $T$, in 1T-TaSe$_2$ flakes with varying thicknesses down to monolayer. The metallic (insulating) behavior of bulk (monolayer) 1T-TaSe$_2$ is consistent with previous reports[34,43]. Remarkably, a metal-to-insulator transition (MIT) occurs at a critical thickness of seven layers. $R_\square$ at the critical thickness is on the order of quantum resistance $h/2e^2$, and stays almost constant over the entire temperature range. The behaviour signifies a quantum phase transition driven by dimensionality reduction. Here $h$ is the plank constant and $e$ the charge of an electron.

The dimensionality-driven MIT observed in transport, however, does not translate to spectroscopic transition in scanning tunneling microscopy and spectroscopy (STM/STS) measurements. Figure 2d displays the differential conductance ($dI/dV$) spectra, which is proportional to energy-resolved local density of states (LDOS), on 1T-TaSe$_2$ flakes with varying number of layers. The spectra were obtained at $T = 4.3$ K on atomically-clean surfaces, as exemplified in Fig. 2c, with an experimental setup sketched in Fig. 2a. Surprisingly, the bulk crystal and all few-layer specimens exhibit a spectral gap of the same size at the Fermi level $E_F$, bracketed by two pronounced peaks that have been referred to as the Hubbard bands[32,37,44]; no clear transition was visible in the spectra of few-layer samples. (The mono-layer spectrum differs from those of thicker samples, probably due to coupling with the conductive Au substrate; Supplementary Fig. 3b). We note that the few-layer spectra seem to be independent of CDW stacking order. Different stacking orders yield identical spectra in few-layer samples (Fig. 2e-g).

The seemingly contradictive results from transport and STS measurements prompt us to examine in detail the electronic structure of bulk 1T-TaSe$_2$ with ARPES. Figure 3a displays the ARPES spectrum along the Γ-K direction acquired with 86 eV $p$-polarized photons at $T = 20$ K. The spectrum reproduces main features of bulk electronic structure reported previously: Se $4p$ band, CDW band folding and a flat band that has been identified as the lower Hubbard band[45]. Surprisingly, however, we observe an additional, dispersive band that crosses the Fermi level, forming a well-defined Fermi surface centered at the Γ point (Fig. 3h). This metallic band is 3D in nature, i.e., the band disperses strongly also in $k_z$, in stark contrast to the 2D Se $4p$ band and the flat band[46] (Supplementary Fig. 6a, b). In particular, the band switches from electron-like (Fig. 3a) to hole-like, and back to electron-like (Supplementary Fig. 5a, b), as $k_z$ traverses the entire Brillouin zone from Γ to A, and back to Γ. The strong $k_z$ dispersion is corroborated by a Fermi surface that is periodic in $k_z$ (Fig. 3c); an observed periodicity of $2\pi/c$ ($c$ is the out-of-plane lattice constant) further indicates the absence of superstructures, such as dimerization[39,42], in the out-of-plane direction.

The metallic band resolves the mystery of metallicity in bulk 1T-TaSe$_2$. But the contradiction between the metallic band and the gapped tunneling spectrum on the same bulk sample remains. A consistent picture emerges once we consider the fact that ARPES has a penetration depth of approximately 1 nm (~ 2 layers; ref. [47]), whereas STM signal is dominated by the surface layer. The ARPES is, therefore, able to access 1T-TaSe$_2$ bulk, in addition to the surface layer which is probed by both ARPES and STM. Our results indicate that 1T-TaSe$_2$ is a bulk metal covered with an insulating surface.

These findings raise a fundamental question: what is the nature of the insulating state? Our experimental evidence indicates that strong correlation is responsible for generating the spectral gap at Fermi level. Key insights come from nano-ARPES measurements of atomically thin 1T-TaSe$_2$ flakes. Figure 3d-f display the ARPES spectra along the Γ-K direction of 1T-TaSe$_2$ flakes of varying thickness down to bilayer (see Methods for details of the measurements). The spectra again reproduce familiar features of the bulk, i.e., Se $4p$ band, CDW band folding and the flat band, all of which do not vary appreciably with sample thickness. The newly discovered metallic band, however, gradually disappears, giving way to an energy gap bounded by the flat band, as the sample becomes thinner (Fig. 3d-f). The same transition is reflected in the vanishing Fermi surfaces of the flakes (Fig. 3i-k); the absence of the metallic band around the A point in a trilayer indicates that the Fermi surface is gapped globally in the 2D limit (Supplementary Fig. 7). The vanishing metallic band corroborates the MIT that we observed in transport measurements of few-layer 1T-TaSe$_2$. More importantly, the way that the band disappears (and the spectral gap develops) reveals important clues on the nature of the spectral gap. There are two main points to notice. First, no energy shift is detected in either the dispersive metallic band or the flat band in the thin flake samples. This observation rules out band bending effect[48] as a cause of the spectral gap.

Second, specimens with even and odd number of layers all exhibit the same spectral gap (Fig. 2d, 3e and 3f). Meanwhile, a constant spectral gap, as opposed to alternating large and small gaps, was observed in the tunneling spectroscopy performed on 1T-TaSe$_2$ terraces with various layer configurations (Fig. 2e-g). These observations, combined with absence of out-of-plane superstructures in bulk 1T-TaSe$_2$ flakes, rules out layer dimerization[39] as the origin of the gap in 1T-TaSe$_2$. Other possibilities of a single-particle gap, such as quantum confinement and weak localization[49,50], are similarly excluded. In fact, we find that no known single-particle mechanism consistently explains the observed insulating state on the surface of metallic 1T-TaSe$_2$ bulk. On the other hand, once electron correlation is considered, the measured spectral gap is naturally explained by a Mott gap. Because 1T-TaSe$_2$ has a cluster CDW superlattice that is exactly half-filled, initially itinerant electrons at material surface (or in few-layers) will make the transition to a Mott insulating state, once the onsite Coulomb energy dominates over the kinetic energy of the electrons.

The question next arises as to what drives the metal to Mott insulator transition both at bulk surface and in few-layer 1T-TaSe$_2$. Prior to answering this question, we first point out that all main features of the ARPES spectra are captured by our density functional theory calculations with self-consistently obtained onsite Hubbard $U$ and inter-site Hubbard $V$ interactions[51,52] (DFT+$U$+$V$; see Methods). Specifically, calculations of the total energy indicate that the energetically most favorable stacking orders are $\pm 2\mathbf{a} + \mathbf{c}$, which yield identical DFT+$U$+$V$ band structures (Supplementary Table 1). The DFT+$U$+$V$ calculations reproduce the dispersive metallic band in the bulk and the gapped spectrum in the bilayer, which agree well with experimental observations (Fig. 3b and 3g). The calculations further reveal that both the dispersive band and the flat band derive from the $d_{z^2}$ orbital of the central Ta atom in each star-of-David cluster (Supplementary Fig. 9). Even though the intra-layer hoping of the $d_{z^2}$ electrons in the CDW phases are strongly suppressed because of the relatively large lateral dimension of the star-of-David cluster, strong inter-layer hoping remains between the pancake-shaped clusters. The inter-layer hoping gives rise to a half-filled dispersive band that hosts itinerant electrons, whose Mott localization turns the material into a Mott insulator.

This revelation is strongly supported by polarized ARPES measurements performed on the surface of 1T-TaSe$_2$ bulk. Figure 4a illustrates our experimental setup of polarized ARPES, where photons impinge on the sample surface with either *p* (in the incident plane) or *s* (perpendicular to the incident plane) polarization. Symmetry dictates that electronic states with odd (even) parity with respect to the mirror plane can be observed with *p* (*s*) polarized photons[53,54]. In addition, the *p*-polarized photons have an out-of-plane (*z*) component, which leads to a large ARPES cross section for states with pronounced out-of-plane character such as the Ta $d_{z^2}$ orbital. The *s*-polarized photons, on the other hand, lie completely in the *x*-*y* plane, and is therefore more sensitive to orbitals with in-plane characters[55,56]. We observe that both the flat band and the dispersive metallic band are visible with *p*-polarized photons, but are absent with *s*-polarized photons (Fig. 4b, d). This observation confirms our calculation that the two bands originate from Ta $d_{z^2}$ orbitals.

We are now poised to explain the underlying mechanism that drives the metal to Mott insulator transition. Physics at the Mott transition is essentially governed by the relative strength of kinetic energy of electrons (characterized by $W$) over onsite Coulomb energy $U$, i.e. $W/U$. Here $U$ represents the bare Coulomb energy cost when an additional electron hops into a star-of-David cluster in 1T-TaSe$_2$. The system is expected to turn into a Mott insulator when $W/U$ falls below a critical value[57-59]. We first rule out the hypothesis that the reduced dielectric screening in few-layer 1T-TaSe$_2$ enhances $U$, and drives the Mott transition. Because the size of the spectral gap seen in STS and ARPES provides a direct measure of Coulomb energy in the Mott insulating state, we are able to extract $U$ as a function of sample thickness (Fig. 2d, 3d-f and Supplementary Fig. 11). We discover that $U$ remains nearly constant as the sample approaches bilayer thickness; the same spectral gap is observed on 1T-TaSe$_2$ bulk surface. We have also calculated self-consistently the onsite Coulomb energy of the center Ta $d$-orbitals, $\widetilde{U}$, with varying number of layers (Supplementary Fig. 11). The thickness-independent $\widetilde{U}$ obtained from the calculations supports our experimental observations, and further rules out Coulomb energy as a driving force of the Mott transition.

Our first-principles calculations, however, reveal a significantly suppressed bandwidth $W$ in atomically thin 1T-TaSe$_2$. Figure 5 displays the width of the dispersive metallic band calculated in a nonmagnetic configuration before electron correlation is added. The bulk band width of ~ 300 meV gradually narrows down to ~ 220 meV as a result of reduced average inter-layer coordination number for the star-of-David clusters, as the sample thickness approaches bilayer. Consequently, the ratio $W/U$ monotonically increases from 0.52 (in bilayer) to 0.67 (in bulk), a range that coincides with theoretically predicted $W/U$ range (0.56 to 0.75) where metal to Mott insulator transition takes place (Fig. 5; ref. [57–59]). It now becomes clear that band narrowing under reduced dimensionality drives the Mott transition in 1T-TaSe$_2$. The dimensionality driven Mott transition also naturally explains the Mott insulating surface layer on bulk 1T-TaSe$_2$: reduced average coordination number at the bulk surface causes a similar dimensionality crossover, inducing a 2D Mott insulating state at the surface. Finally, we note that even though we only considered coordination number reduction, slight lattice distortions such as swelling in $c$-axis (seen in 1T-TaS$_2$ thin flakes; ref. [60]) may potentially play a role in suppressing the bandwidth in few layer samples. Such distortions are beyond the detection limit in the present study, and future work is needed to delineate their contribution to the Mott transition.

To summarize, we discover an intriguing metal to Mott insulator transition that is driven by dimensionality reduction. Specifically, we provide unequivocal evidence that the insulating phase in few-layer 1T-TaSe$_2$ and on bulk 1T-TaSe$_2$ surface is a strongly correlated Mott insulator, which evolves from a 3D metallic band that we observe for the first time in bulk 1T-TaSe$_2$. Below a critical thickness of seven layers, reduced average coordination number effectively suppresses the bandwidth (i.e., the kinetic energy of the electron), and turns the initially bulk metal into a 2D correlated Mott insulator. A similar dimensionality crossover induces a Mott insulating layer on the bulk surface, resolving the long-standing dichotomy between the metallic bulk and insulating surface in 1T-TaSe$_2$. Our work establishes few-layer 1T-TaSe$_2$ as a fully tunable material system with relatively high correlation energy on the scale of 0.5 eV, where various emergent correlated quantum phenomena such as quantum spin liquid and high-temperature superconductivity may be investigated in the extreme 2D limit.

## Methods

**Growth and characterization of bulk 1T-TaSe$_2$ crystal.** High-quality 1T-TaSe$_2$ single crystals were grown by the chemical vapour transport (CVT) method. High-purity elements were stoichiometrically mixed and sealed under vacuum ($3 \times 10^{-5}$ mbar) in a quartz ampoule with a very small quantity of iodine as the transport agent (7‰ mass fraction). The ampoules were then placed horizontally into a tube furnace. The hot and cold ends of the tube were kept at 1050℃ and 950℃, respectively, during growth. To achieve high crystal quality and to minimize defect density, the ampoules were kept at growth temperature for one month. The ampoules were then rapidly quenched in cold saturated salt-water ($\sim -15$℃) at the end of the growth. The as-grown 1T-TaSe$_2$ single crystals typically have a dimension of 4 mm × 4 mm × 0.2 mm (Supplementary Fig. 1a). Supplementary Fig. 1b displays the X-ray diffraction pattern of a typical crystal, which indicates the good crystallinity of bulk 1T-TaSe$_2$.

**STM and STS measurements on 1T-TaSe$_2$ thin flakes.** We used oxidized Si wafers covered with 2 nm of Cr and 3 nm of Au as substrates, on which we prepared atomically clean thin 1T-TaSe$_2$ flakes for STM measurements in two steps. We first pressed pieces of bulk 1T-TaSe$_2$ crystal, fixed on a piece of vacuum-compatible tape, onto the substrate in a glovebox. The water and oxygen level in the glovebox was kept below 0.1 ppm during the process. The substrate, along with the 1T-TaSe$_2$ crystal and the tape, was subsequently transferred into ultra-high vacuum (UHV). We then peeled away the tape under pressures below $5 \times 10^{-9}$ mbar, leaving atomically clean few-layer 1T-TaSe$_2$ on the substrate. The number of layers is determined from optical contrast combined with STM measurements of the step edges on the flakes. Our STM and STS measurements were performed in a Createc low-temperature STM at $T = 4.3$ K in UHV environment with base pressure blow $2 \times 10^{-10}$ mbar. We used electrochemically etched polycrystalline tungsten tips in all our STM measurements. All tips were calibrated by tunnelling differential conductance ($dI/dV$) measurements on Au (111) surface before STM measurements on 1T-TaSe$_2$. $dI/dV$ spectra were obtained through lock-in detection with an excitation wiggle voltage $V_{\text{r.m.s.}}$ ranging from 5 to 10 mV at frequency $f = 444$ Hz.

**ARPES Measurements.** ARPES measurements on freshly cleaved bulk 1T-TaSe$_2$ crystals were performed at 03U and 09U beamlines of Shanghai Synchrotron Radiation Facility (SSRF; the polarized ARPES was performed at 09U beamline in particular). Nano-ARPES measurements on 1T-TaSe$_2$ thin flakes were conducted at 07U beamline of SSRF. All end stations are equipped with Scienta Omicron DA30 electron analyzers, which have an angular resolution better than 0.2°. The energy resolution is 30 meV at 03U and 09U, and 35 meV at 07U. All ARPES data were acquired at temperatures below the CDW transition temperature of 473 K (20 K at 03U and 09U, and 40 K at 07U) under ultrahigh vacuums better than $8.0 \times 10^{-11}$ Torr. Some of the preliminary ARPES data was collected at the BL 10.0.1 of the Advanced Light Source.

The Nano-ARPES has a spatial resolution of 400 nm, which enables us to probe the local ARPES intensity within a sample (the same type of thin flake sample used in STM measurements; typical size ~ 50 μm × 50 μm). Locating the thin flakes in the nano-ARPES setup, however, poses a challenge. To this end, we marked the position of the sample area with a sharp tip before loading the samples into the measurement setup. We then scanned the sample area and acquired a map of the ARPES intensity at a photoelectron energy of 86 eV. The large ARPES intensity from Au substrate gives a clear contrast between the sample and the substrate. The intensity map resolves the profile of the flakes, which allows us to perform ARPES measurements on areas with uniform thickness.

**First-principles calculations of the electronic structure.** First-principles density-functional theory calculations (referred to as DFT+$U$+$V$) were performed with Quantum Espresso package and Garrity-Bennett-Rabe-Vanderbilt ultrasoft pseudopotentials[61]. The kinetic energy cutoff for charge density was set to 200 Ry, and $5 \times 5 \times 5$ k-point mesh was used for self-consistent calculations of $\sqrt{13}$ by $\sqrt{13}$ by 2 star-of-David CDW structures for all stacking orders under investigation. The CDW structures were fully relaxed with the rev-vdW-DF2 functional[62].

We adopted the newly developed pseudo-hybrid density functionals for self-consistently obtained extended Hubbard interaction to calculate the electronic structure[51,52]. This method has been shown to describe the quasiparticle energy bands of various solids such as semimetal, semiconductor, ionic insulator and Mott insulator with an accuracy comparable to the $GW$ approximation[51]. The self-consistently computed Hubbard parameters are consistent with those obtained with other methods[51,52]. The method also describes the structural properties for semiconducting silicon, diamond and germanium crystals as well as Mott insulating NiO and MnO very well[52,63]. In particular, the computed phonon dispersion for Mott insulators NiO and MnO agree well with experimental results and dynamical mean field theory calculations[63]. These examples have demonstrated the capability of in calculating the structural property and the electronic structure, as well as the coupling between the two, in correlated systems such as 1T-TaSe$_2$.

The onsite Hubbard $\widetilde{U}$ and inter-site Hubbard $V$ parameters are self-consistently obtained from undistorted $1 \times 1$ bulk structure. The obtained $\widetilde{U}$ value for Ta $d$ orbital is 1.1 eV, and the $V$ value between Ta $d$ and Se $p$ orbitals vary from 1.7 to 2.2 eV depending on their bond direction for a bulk case. As the number of layers increases from monolayer to bulk, we found that the self-consistent $\widetilde{U}$ value for the center Ta atom of the David star varies from 1.304 eV to 1.1904 eV (Supplementary Fig. 11). For direct comparison with ARPES spectra, the band structure of the CDW superlattice were unfolded into the Brillouin zone of undistorted $1 \times 1$ structure. The unfolding was performed using the BandUP code[64–66]. The band broadening was simulated with an imaginary part of self-energy of the form $a\omega^2 + b$, where $a = 1.0 \text{ eV}^{-1}$, $b = 0.1 \text{ eV}$, and ω is an energy difference between the Fermi energy and band energy.

initio band structures. *Comput. Phys. Commun.* **272**, 108226 (2022).

**Acknowledgements**

We thank Cenyao Tang, Liguo Ma, Hongya Wang, Yujun Deng, Di Yue and Zhiwei Huang for helpful discussions. We also thank Yiwei Li, Hanbo Xiao, Han Gao for help with nano-ARPES measurements. Part of the sample fabrication was conducted at Nano-fabrication Laboratory at Fudan University. N.T., S.G., Y.Yu. and Y.Z. acknowledge financial support from National Key R&D Program of China (Grant No. 2018YFA0305600), Strategic Priority Research Program of Chinese Academy of Sciences (Grant No. XDB30000000), and Shanghai Municipal Science and Technology Commission (Grant No. 2019SHZDZX01). W.R. acknowledges support from the young scientist project of MOE innovation platform. Z.H. and D.S. acknowledge support from National Science Foundation of China (Grant No. U2032208). Part of this research used 03U Beamline of the Shanghai Synchrotron Radiation Facility, which is supported by ME$^2$ project under Grant No. 11227902 from National Natural Science Foundation of China. Y.-W.S. acknowledges support from NRF of Korea (Grant No. 2017R1A5A1014862, SRC program: vdWMRC center) and KIAS individual Grant (No. cG031509). M.K. and B.G.J. acknowledge support from KIAS individual Grants (No. CG083501 and QP081301). M.W. and Z.L. acknowledge support from the National Natural Science Foundation of China (Grants No. 12274298) and Double First-Class Initiative Fund of ShanghaiTech University. S.-K.M. acknowledges support from the Office of Basic Energy Sciences, the US Department of Energy under Contrast No. DE-AC02-05CH11231. X.L. and Y.P.S. acknowledge support from National Key R&D Program of China (Grant No. 2021YFA1600201 and the National Natural Science Foundation of China (Grants No. 11674326, 11874357, 12274412, U1832141, U1932217, U2032215 ).

**Author Contributions**

Y.Z., W.R., D.S and Y.-W.S. supervised the project. N.T., Y.-J.Y., J.G., X.L., Y.P.S., D.-L.F. and X.C. grew bulk 1T-TaSe$_2$ crystals. N.T. and Y.Y. fabricated 1T-TaSe$_2$ thin flake samples. N.T. performed electronic transport and STM measurements. N.T., Z.H., S.G., J.H., M.W., Z.L., S.-K.M. and D.S. performed ARPES measurements. B.G.J., M.K. and Y.-W.S carried out first-principles calculations. N.T., W.R and Y.Z.

wrote the manuscript with input from all authors.

## Competing interests

The authors declare no competing interests.

## Materials & correspondence

Correspondence and requests for materials should be addressed to Y.Z., W.R., D.S. or Y.-W.S.

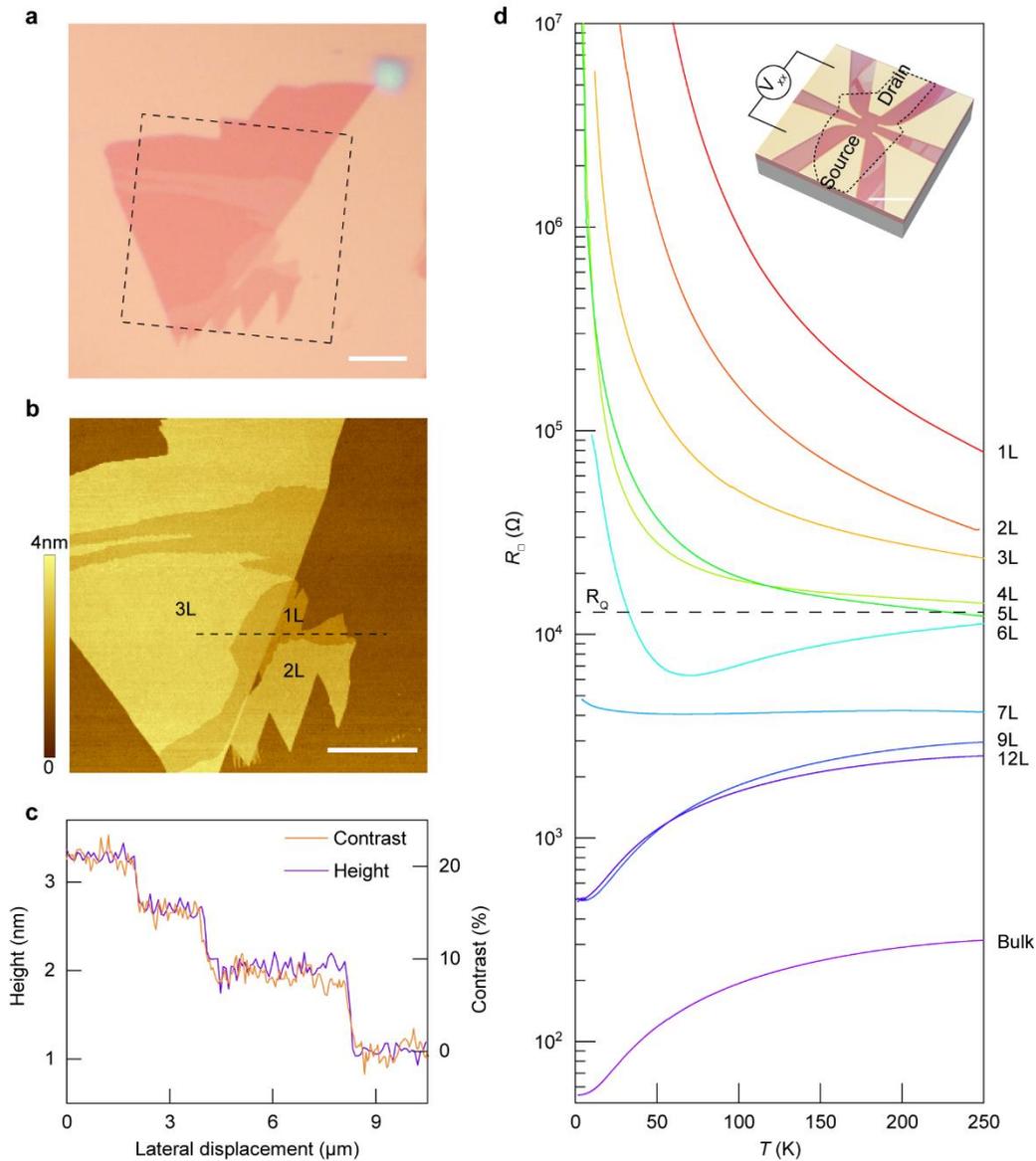

**Fig. 1 | Metal to insulator transition in atomically thin 1T-TaSe$_2$ flakes of varying thicknesses. a**, Optical image of a typical 1T-TaSe$_2$ thin flake mechanically exfoliated on the substate. The substrate is Si wafer covered with 285-nm-thick SiO$_2$. Scale bar, 5µm. **b**, Atomic force microscopy (AFM) image of the area marked by the dashed square shown in **a**. The number of layers, determined from AFM measurement of the sample thickness, is marked on various parts of the flakes ('1L', '2L' and '3L' denote monolayer, bilayer and trilayer, respectively). Scale bar, 5 µm. **c**, Cross-sectional AFM height profile along the line cut marked by the black dashed line in **b**. Superimposed on top is the optical contrast profile extracted from the optical image in **a** along a same line cut. The good agreement between the two profiles indicates that both methods— optical contrast and AFM—give accurate measurements of the sample thickness. **d**,

Temperature-dependent resistivity of 1T-TaSe$_2$ flakes with varying number of layers. A metal to insulator transition is clearly visible at the critical thickness of seven layers. Data from bulk 1T-TaSe$_2$ is also shown as a reference. The broken line denotes the value of quantum resistance $R_Q = h/2e^2$. Inset: Optical image of a bilayer 1T-TaSe$_2$ device with schematic measurement setup sketched on top.

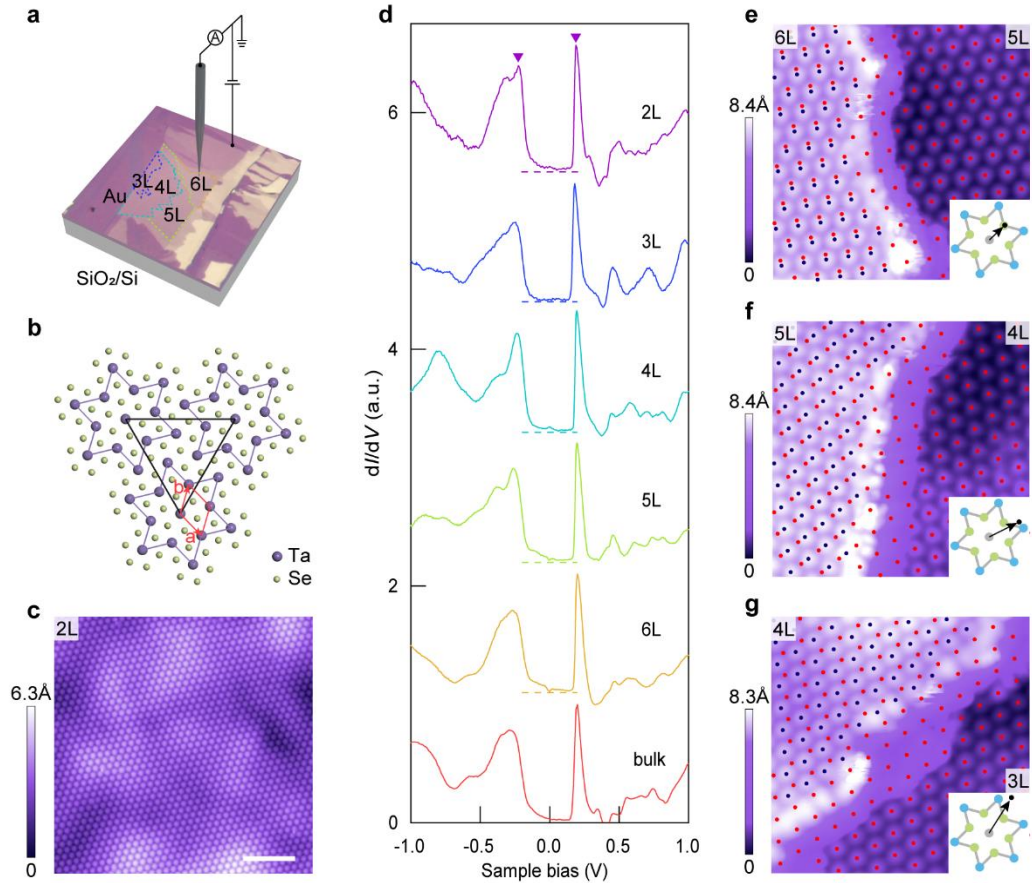

**Fig. 2 | Scanning tunnelling microscopy and spectroscopy of few-layer 1T-TaSe$_2$ in the star-of-David CDW phase. a**, Few-layer 1T-TaSe$_2$ flake exfoliated on Au-covered SiO$_2$ substrate probed by STM (STM tip is sketched). Number of layers determined from optical contrast and STM measurements is marked on different parts of the flake. **b**, Top view of the monolayer 1T-TaSe$_2$ crystal structure. Every 13 Ta atoms form a star-of-David cluster (outlined in purple). The CDW superlattice is formed by the star-of-David clusters arranged in a triangular lattice. ***a*** and ***b*** denote the in-plane unit vectors of the atomic lattice. **c**, Constant-current STM topography of a bilayer 1T-TaSe$_2$ ($V_b = 0.3$ V, $I_t = 50$ pA). The large-scale ripples reflect corrugations of the substrate. Scale bar, 10 nm. **d**, Differential conductance spectra acquired on 1T-TaSe$_2$ flakes with varying number of layers ($V_b = 1$ V, $I_t = 200$ pA, $V_{r.m.s.} = 5$ mV). Spectra are vertically displaced for clarity. Broken lines mark the zero of each curve. All spectra show a spectral gap (bounded by the two peaks marked by the two triangles on each curve), which does not vary with sample thickness. **e-g**, Constant-current STM topography ($V_b = 0.3$ V, $I_t = 50$ pA) of a 6L to 5L step edge, a 5L to 4L step edge, and a 4L to 3L step edge. Centers of the star-of-David clusters are marked by blue (red) dot

array on the upper (lower) terraces in each panel. By extrapolating the position of the array on the lower terrace to the upper terraces, we extract the stacking order between the two terraces, which is depicted in the inset of each panel. Insets: Schematic illustration of the stacking order determined from STM topographies. The black dot and arrow indicate the lattice shift of the upper terrace relative to the lower terrace. All STM and tunneling spectroscopy data were taken at $T = 4.3$ K.

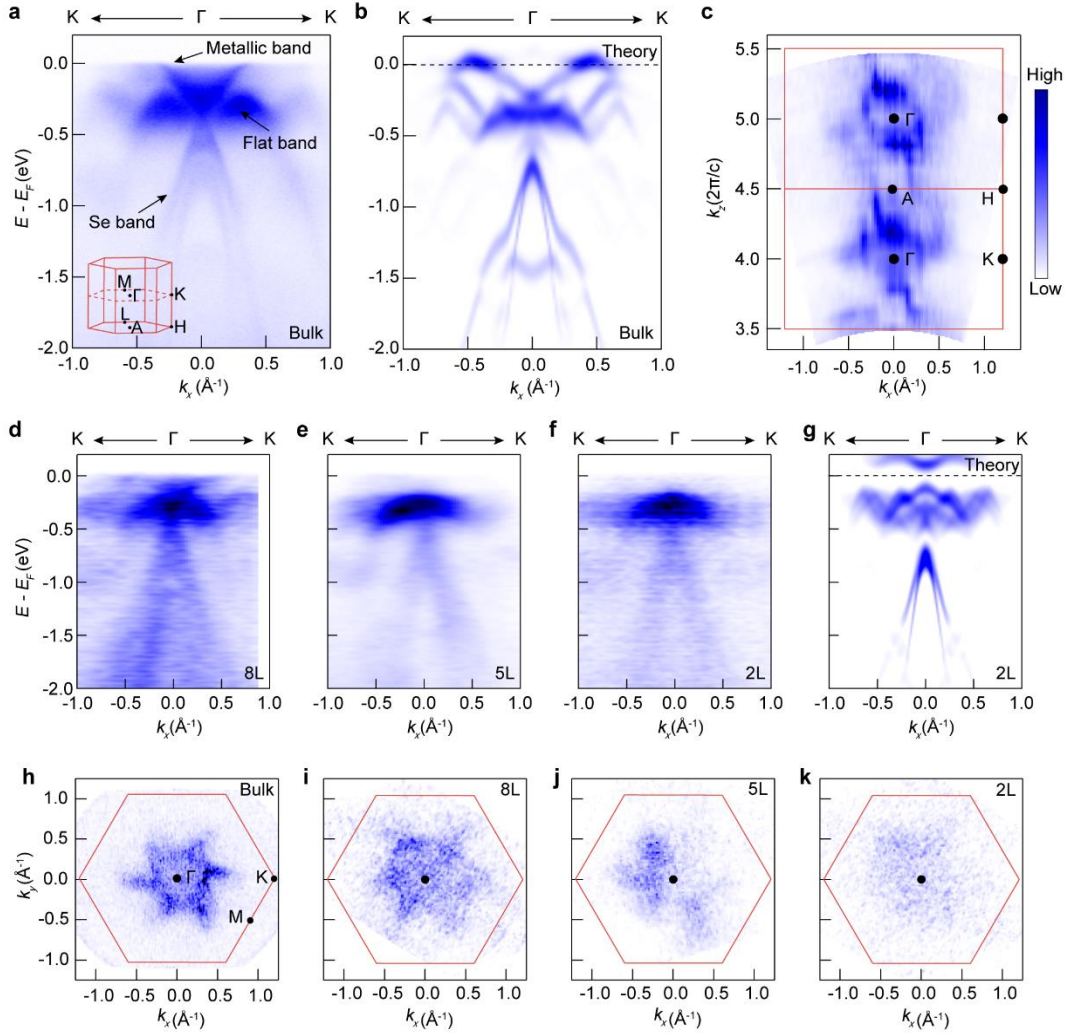

**Fig. 3 | Electronic structure of bulk and few-layer 1T-TaSe₂ probed by ARPES. a**, ARPES spectrum of bulk 1T-TaSe$_2$ acquired with 86 eV *p*-polarized photons at $T = 20$ K along Γ-K direction of the undistorted (i.e., no CDW) atomic lattice Brillouin zone. The full Brillouin zone is sketched in the inset. Apart from the previously reported flat band and Se band (in, e.g., ref. [46]), we discover an additional dispersive band that crosses the Fermi level at $k_x \sim 0.3$ Å$^{-1}$. **b**, DFT+*U*+*V* band structure of bulk 1T-TaSe$_2$. The calculation is performed on the star-of-David CDW superlattice with $2\mathbf{a} + \mathbf{c}$ interlayer stacking order. The calculated band structure was unfolded onto the undistorted atomic lattice Brillouin zone for comparison with ARPES spectrum from **a**. The calculation reproduces all main features of the ARPES spectrum, including the dispersive metallic band. **c**, Spectral weight mapping near Fermi level ([−10 meV, 10 meV]) in the $k_x - k_z$ plane of bulk 1T-TaSe$_2$ taken at $T = 20$ K. Here $k_x$ is defined along the Γ-K direction. A Fermi surface having a period of $2\pi/c$ in $k_z$ is clearly visible. **d-f**, ARPES spectra along the Γ-K direction of undistorted atomic lattice

Brillouin zone of 8-layer (8L; **d**), 5-layer (5L; **e**) and bilayer (2L; **f**) 1T-TaSe$_2$ flakes. Data were taken with 86 eV circularly-polarized photons at $T = 40$ K in a nano-ARPES setup. **g**, DFT+$U$+$V$ band structure of bilayer 1T-TaSe$_2$ with $2\mathbf{a} + \mathbf{c}$ CDW stacking order for comparison with the ARPES spectrum from **f**. **h-k**, Spectral weight mapping acquired with 86 eV photons near Fermi level ([−1 meV, 1 meV]) in the $k_x - k_y$ plane of bulk (**h**), 8L (**i**), 5L (**j**) and 2L (**k**) 1T-TaSe$_2$.

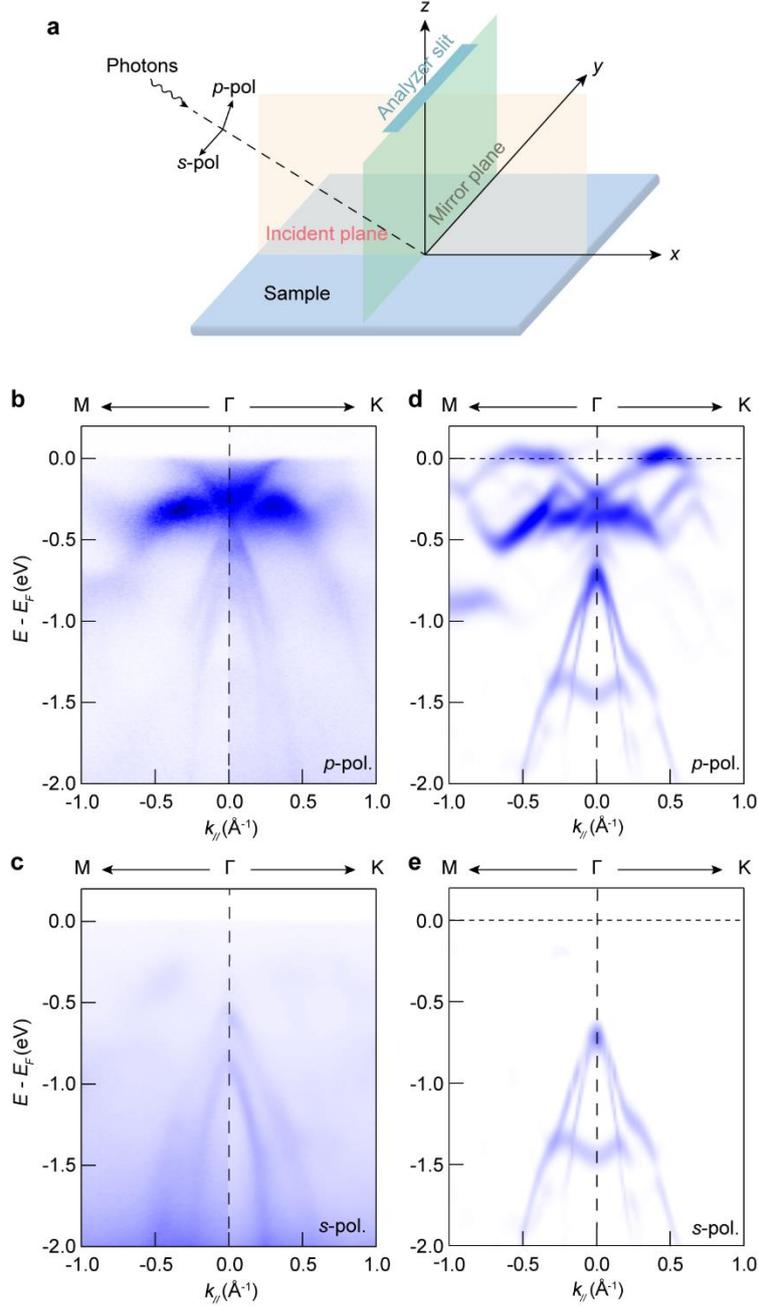

**Fig. 4 | Orbital character of electronic bands in bulk 1T-TaSe$_2$ probed by polarized ARPES. a**, Schematic illustration of our polarized ARPES measurement setup. The analyzer slit is perpendicular to the incident plane (orange). The mirror plane (green) is defined by the analyzer slit direction and the sample surface normal. **b**, **c**, ARPES spectra along the Γ-K and Γ-M of bulk 1T-TaSe$_2$ acquired with 86 eV $p$ (**b**) and $s$ (**c**) polarized light at $T = 20$ K. **d, e,** DFT+$U$+$V$ band structure of bulk 1T-TaSe$_2$ for comparison with polarized ARPES spectra from **b** and **c**, respectively. Panels **d** and **e** depict the calculated band structure with and without Ta $d_{z^2}$ orbital contribution, respectively.

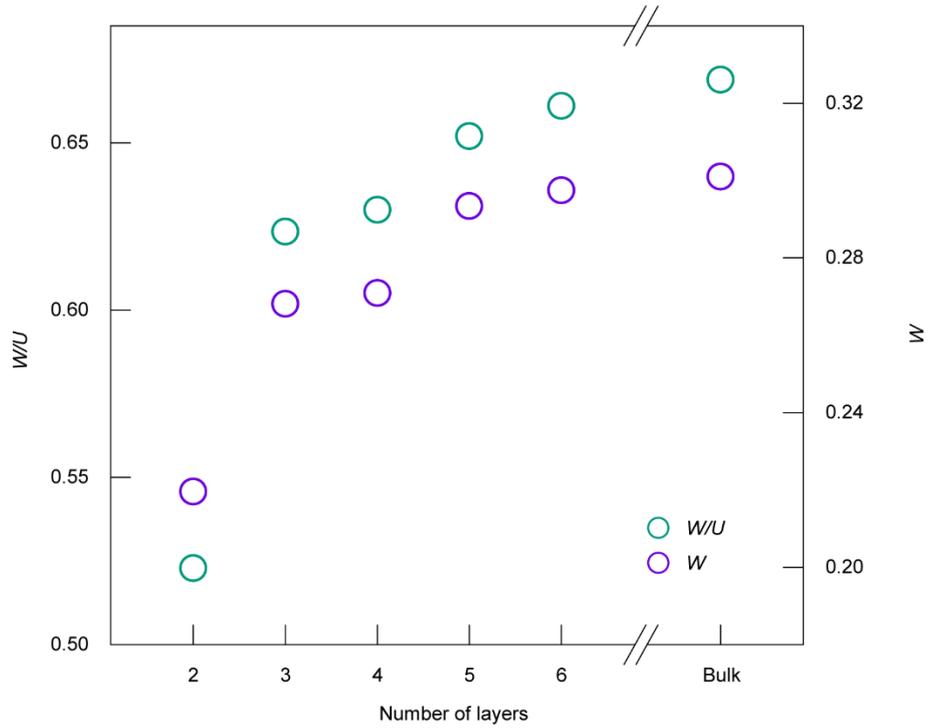

**Fig 5. | Narrowing of the metallic band and metal to Mott insulator transition in few-layer 1T-TaSe$_2$.** The bandwidth of the metallic band, $W$, and the ratio between $W$ and the onsite Coulomb energy, $W/U$, are plotted as functions of the number of layers. Here $W$ is obtained from DFT+$U$+$V$ non-magnetic calculations of the metallic band, and $U$ is extracted from the size of the spectral gap seen in STS. The $W/U$ in few-layer 1T-TaSe$_2$ fall within the range where metal to Mott insulator transition is anticipated (see text).